INFERENCE

# Inferring propagation paths for sparsely observed perturbations on complex networks

Francesco Alessandro Massucci,[1] Jonathan Wheeler,[1]* Raúl Beltrán-Debón,[2] Jorge Joven,[3] Marta Sales-Pardo,[1†‡] Roger Guimerà[1,4†]

In a complex system, perturbations propagate by following paths on the network of interactions among the system's units. In contrast to what happens with the spreading of epidemics, observations of general perturbations are often very sparse in time (there is a single observation of the perturbed system) and in "space" (only a few perturbed and unperturbed units are observed). A major challenge in many areas, from biology to the social sciences, is to infer the propagation paths from observations of the effects of perturbation under these sparsity conditions. We address this problem and show that it is possible to go beyond the usual approach of using the shortest paths connecting the known perturbed nodes. Specifically, we show that a simple and general probabilistic model, which we solved using belief propagation, provides fast and accurate estimates of the probabilities of nodes being perturbed.

## INTRODUCTION

Consider the following situation: An individual ingests a drug, and hours later, a blood test reveals that the concentration of some metabolites has changed with respect to the physiological baseline. The physician trying to interpret the chain of events that led to these changes faces significant challenges. First, there are thousands of metabolites and thousands of biochemical reactions that transform metabolites into one another. Second, the structure of the metabolic network defined by these reactions is highly nontrivial (1–4). Third, the physician ignores the state of most metabolites—only a few are measured by the blood test. Through which reactions did the perturbation spread? Which of the metabolites that cannot be measured have also been perturbed?

The challenge of reconstructing the state of a perturbed system from a single sparse observation is very common in biology and biomedicine, as in the example above or when comparing diseased versus healthy states or treatments versus controls. However, this situation may also arise in very different contexts. For example, consider an organization where a confidential memo has been leaked through email. A few members of the organization are known to have the memo, a few others are known not to have it, and one may know who sent emails to whom (5–7) but not the content of the emails. Which path within the organization did the leakage follow? Who else has the memo?

At an abstract level, both situations above are analogous in two aspects: A perturbation spread through a system whose elements are connected in a known complex network of interactions, and we wish to infer the propagation path from a single observation that provides information about the final state (perturbed or unperturbed) of only a small fraction of the nodes (Fig. 1).

Although this problem is conceptually similar to the relatively well-studied problem of locating the source of a network cascade [for example, of an epidemic outbreak (8–13) or of an information cascade (14)], it is more challenging in practice because (i) only a single observation of the perturbed system is available [as opposed to having observations at multiple times or to having the exact "infection time" of each node (10–13, 15–17), and to having observations of multiple cascades (16)]; (ii) the observation of the system is very sparse for both perturbed and unperturbed nodes [as opposed to situations in which we have complete or almost complete information about, at least, the perturbed nodes (11, 13, 16)]; and (iii) we lack a model for the process through which the perturbation spreads [as opposed to situations in which there is a reasonable model, such as the SIR (susceptible-infected-recovered), for the propagation process (10, 12, 13, 15), let alone reasonable estimates of model parameters (13)].

Despite the importance and ubiquity of the problem of inferring perturbation propagation paths under the aforementioned sparsity conditions, we still lack a systematic approach to addressing it. As a result, it is common to resort to the so-called "network parsimony principle." In the context of network medicine, this principle states that "causal molecular pathways often coincide with the shortest molecular paths between known disease-associated components" (18); more broadly, the parsimony principle assumes that perturbation propagation paths coincide with the shortest paths in the complex network underlying the propagation process (19).

Here, we show that a simple and general probabilistic model, on which we can use belief propagation inference (20), provides accurate estimates of the probabilities of nodes being perturbed. Specifically, we show that our method performs better than potential alternatives for (i) synthetic perturbations in model and real-world networks and (ii) real metabolic perturbation measured in human patients after ingestion of a plant extract.

## RESULTS
### Model

Our goal is to develop an approach that is general enough to be of use for disparate perturbation propagation processes. Therefore, we propose a probabilistic model for the final perturbed state of the system rather than for the propagation process itself. In the model, each node $i$ is in one of two states: $\sigma_i = 1$ if the node has been perturbed during the propagation or $\sigma_i = 0$ if it has not been perturbed. We assume that the probability of the state $\sigma_i$ of each unobserved node depends exclusively on the state $\gamma_i$ of its neighbors

[1]Departament d'Enginyeria Química, Universitat Rovira i Virgili, Tarragona 43007, Catalonia, Spain. [2]Cheminformatics and Nutrition Research Group, Department of Biochemistry and Biotechnology, Universitat Rovira i Virgili, Tarragona 43007, Spain. [3]Unitat de Recerca Biomèdica, Hospital Universitari de Sant Joan, Institut d'Investigació Sanitària Pere Virgili, Universitat Rovira i Virgili, Reus, Catalonia, Spain. [4]Institució Catalana de Recerca i Estudis Avançats (ICREA), Barcelona 08010, Catalonia, Spain.
*Present address: Facultad de Ciencias Exactas y Tecnología, Universidad Nacional de Tucumán, Avda. Independencia 1800, San Miguel de Tucumán, Argentina.
†These authors contributed equally to this work.
‡Corresponding author. Email: marta.sales@urv.cat

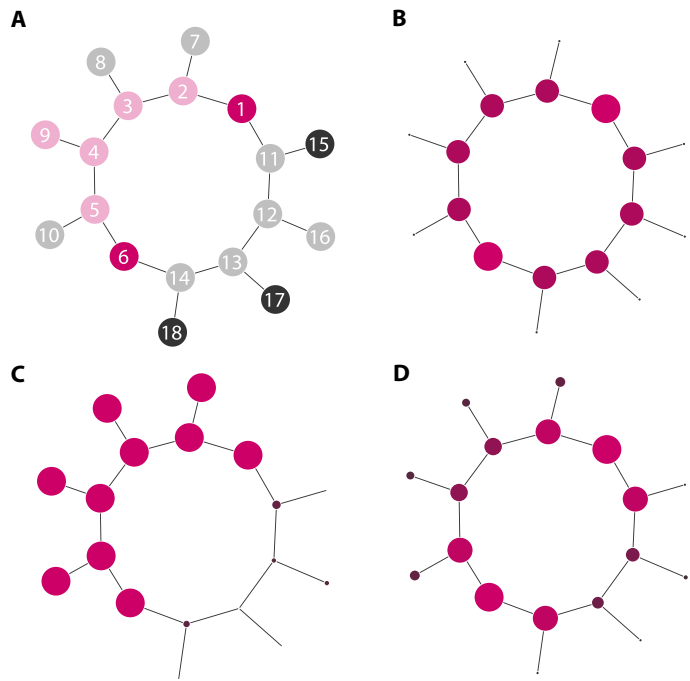

**Fig. 1. Shortest paths, label propagation, and the exposure model for inferring the propagation path of perturbations.** (**A**) A perturbation spread from node 1 to node 6 through nodes 2 to 5, additionally branching out from node 4 to node 9 (perturbed nodes are light red, whereas unperturbed ones are gray). To the observer, the only available information is that nodes 1 and 6 are perturbed and that nodes 15, 17, and 18 are not (observed nodes are dark-colored, whereas unobserved nodes are light-colored). (**B** to **D**) The color and size of the nodes indicate the ranking assigned by the shortest paths approach (B), label propagation (C), and the exposure model (D); larger red nodes are deemed more likely to be perturbed, whereas smaller darker nodes are deemed less likely. (B) By using shortest paths to infer the perturbation propagation path, nodes 2 to 5 and 11 to 14 are all assigned the same probability of being perturbed (because both paths connecting 1 and 6 are equally short), whereas all other nodes are deemed equally unlikely of being perturbed (because no shortest path traverses them). Therefore, by considering shortest paths, all information provided by the observed unperturbed nodes 15, 17, and 18 is ignored. (C) Label propagation assigns the maximum probability of being perturbed not only to nodes in the 2-to-5 path but also to nodes 7 to 10, because no observed node in that region of the network is unperturbed. (D) The exposure model approach exploits the information provided by the observed unperturbed nodes and assigns a higher probability to the 1-to-5 path than to the 11-to-14 path. Additionally, it acknowledges that branching out of this path into nodes 7 to 10 is possible and more likely than branching out into node 16.

$$P(\sigma_i = 1 \mid \gamma_i, \sigma^{\mathcal{O}}) = \begin{cases} 0 & \text{if } \gamma_i = 0 \\ \eta & \text{if } \gamma_i = 1 \end{cases} \qquad (1)$$

where $\gamma_i = 0$ if all neighbors of $i$ are unperturbed and $\gamma_i = 1$ otherwise, and $\eta$ is a parameter that determines how easy it is for the perturbation to propagate. Therefore, the state of a node only depends on whether or not it is exposed to the perturbation; thus, we call the model the exposure model.

From the exposure model, we are ultimately interested in determining the marginal probabilities $P(\sigma_i | \sigma^{\mathcal{O}})$ of the state of each unobserved node, given the state $\sigma^{\mathcal{O}}$ of the observed ones. These can be obtained by marginalizing the joint probability $P(\sigma, \eta | \sigma^{\mathcal{O}})$ over $\eta$ and the state $\sigma$ of all unobserved nodes other than $i$

$$P(\sigma_i | \sigma^{\mathcal{O}}) = \int_0^1 d\eta \sum_{\sigma_{j \neq i} \in \{0,1\}} P(\sigma, \eta | \sigma^{\mathcal{O}}) \qquad (2)$$

Estimating this integral using Markov chain Monte Carlo involves exploring a $2^M$-dimensional space (with $M$ as the number of unobserved nodes) for each value of $\eta$, which becomes impractical for large networks. To circumvent this limitation, we consider an approximation based on belief propagation (*20*) (see Materials and Methods), which scales well with size and, most importantly, yields accurate results.

To start, we relate $P(\sigma_i | \sigma^{\mathcal{O}})$ to the posterior marginal $P(\sigma_{\partial i} | \sigma^{\mathcal{O}})$ of the neighbors of $i$ using standard Bayesian methods (see Materials and Methods). Then, we make the following two approximations: First, we approximate $P(\sigma_{\partial i} | \sigma^{\mathcal{O}})$ by computing it in the absence of $i$; second, we assume that the system is tree-like, such that there are no (short) paths connecting the neighbors of $i$ (other than through $i$ itself). As a consequence, their joint marginal in the absence of $i$ factorizes. Albeit the effect of loops on the results produced by belief propagation is far from trivial to quantify, these standard approximations have been observed in several cases to not seriously undermine the validity of the results, even on complex networks (*12*, *21*, *22*).

Within these assumptions, we can finally write the probability for the state of node $i$

$$P(\sigma_i | \sigma^{\mathcal{O}}) = \sum_{\sigma_{\partial i}} P(\sigma_i | \gamma_i, \sigma^{\mathcal{O}}) \prod_{j \in \partial i} P^{(i)}(\sigma_j | \sigma^{\mathcal{O}}) \qquad (3)$$

where the sum is a trace over the state of the neighbors of $i$, and $P^{(i)}(\sigma_j | \sigma^{\mathcal{O}})$ is the marginal of node $j$ in the system without node $i$. Note that, for the sake of clarity, we have momentarily dropped the reference to $\eta$.

Within our approach, the quantities $P^{(i)}(\sigma_j | \sigma^{\mathcal{O}})$ for each node pair $i$ and $j$ can be computed recursively (see Materials and Methods). In particular, because $\sigma_i$ is a binary variable, we only need one parameter to describe its probability marginals; for notation compactness, we thus define the "beliefs" $\psi_i := P(\sigma_i = 0 | \sigma^{\mathcal{O}})$ (related to Eq. 3) and the "messages" $\psi_i^{(j)} := P^{(j)}(\sigma_i = 0 | \sigma^{\mathcal{O}})$ (see Materials and Methods). For each fixed value of $\eta$, we can compute messages and beliefs using the following relations

$$\psi_i^{(k)} = \left[\prod_{j \in \partial i \setminus k} \psi_j^{(i)}\right] + (1-\eta)\left[1 - \prod_{j \in \partial i \setminus k} \psi_j^{(i)}\right] \qquad (4)$$

$$\psi_i = \left[\prod_{j \in \partial i} \psi_j^{(i)}\right] + (1-\eta)\left[1 - \prod_{j \in \partial i} \psi_j^{(i)}\right] \qquad (5)$$

The belief propagation terminology may be understood as follows: $\psi_i^{(k)}$ is a message that $i$ delivers to $k$ about its own state computed in the absence of $k$, that is, with no knowledge of the state of $k$ and the remaining neighbors of that node. The belief $\psi_i$ is instead the probability that node $i$ is perturbed, as computed in the full system.

Finally, we need to compute the integral over $\eta$ in Eq. 2. We choose to evaluate the integral using a saddle point approximation, that is, by computing the integral at the value $\eta^\star$, which maximizes the joint posterior $P(\sigma, \eta | \sigma^{\mathcal{O}})$. Expressing this posterior in an approximated factorized form, explicit differentiation shows that $\eta^\star$ is given by the fraction of nodes having perturbed neighbors that are also perturbed (see Materials and Methods). Hence, within our approach, we can write

$\eta^*$ in terms of the marginals, Eqs. 4 and 5 (see Materials and Methods), as

$$\eta^* = \frac{\sum_i (1-\psi_i)\left(1 - \prod_{j \in \partial i} \psi_j^{(i)}\right)}{\sum_i \left(1 - \prod_{j \in \partial i} \psi_j^{(i)}\right)} \quad (6)$$

Through the iteration of Eqs. 4 to 6, we can obtain estimates for the probability of each unobserved node being perturbed.

In the following sections, we validate our approach on real networks for both synthetic and real perturbations. We benchmark our method both with shortest paths (the parsimony principle, which is a more conventional tool in the network science community) and with a technique derived from the context of machine learning, that is, label propagation (23, 24). This method was not originally devised to deal with the problem at stake, but we have adapted it to this study by recasting the perturbation inference problem as a classification task on partially labeled graphs (23, 25). In contrast to most popular methods for label learning that have limited applicability to our problem [for example, Laplacian regularization methods, which have limited applicability to large networks (25, 26)], label propagation is, in principle, easily adaptable to the present case. We also compare our approach to the performance of a $k$–nearest neighbor classifier by adapting yet another machine learning problem, namely, the optimal active search, of which our perturbation inference may be seen as the one-step lookahead case on a graph (27) (see the Supplementary Materials).

**Validation on synthetic perturbations**

We first test our method by generating synthetic perturbations, simulating sparse observations of the perturbed system, and inferring the state of the unobserved nodes. To generate the perturbation, we use two different variants of the susceptible-infected (SI) model (28), wherein we attempt to perturb each node only once. In the first case, the probability of propagation $c$ is homogeneous. To generate the perturbation, we select a root node $\mathcal{R}$ at random, we set it as perturbed, and we propagate the perturbation to each of its neighbors with probability $c$; in turn, each of the perturbed neighbors forwards the perturbation to its neighbors. In the second case, we consider an SI model with a heterogeneous probability of propagation $c$; specifically, $c$ is a random number uniformly distributed in the range $\Delta = [l, u]$, where $0 \le l \le u \le 1$ (more details are given in the Supplementary Materials). In both cases, we do not try to perturb a given node more than once if the first attempt was unsuccessful, so that the perturbation stops when there are no newly perturbed nodes. Note that the above processes are arbitrary and completely independent of our inference protocol based on the exposure model.

After generating the perturbation, we observe the state of a set $\mathcal{O}$ of nodes and hide the state of the others. We then apply our algorithm to infer the state of all nodes not belonging to $\mathcal{O}$. We evaluate the performance of our approach by calculating the frequency with which, given a pair of unobserved nodes whose states are perturbed and unperturbed, respectively, the algorithm identifies the perturbed node as being more likely to be perturbed than the unperturbed one. This is equivalent to computing the so-called area under the receiver operating characteristic curve (AUC) (29) (see Materials and Methods). We carry out this protocol for different sizes of $\mathcal{O}$, by varying the value of $c$ and of the range width $\delta_c = u - l$ for the homogeneous and heterogeneous SI models, respectively.

To test the accuracy of our method at inferring these synthetic perturbations, we repeat the two abovementioned perturbation protocols on the giant component of four real networks with different average degrees $\langle k \rangle$ and different numbers of nodes $N$, where perturbation processes, such as the one we are considering, are likely to take place. These networks are (i) the metabolic network of the bacterium *Escherichia coli* ($\langle k \rangle$ = 2.8, $N$ = 507) (4, 30, 31), where the alteration of metabolite concentrations can result from the intake of a drug; (ii) the global air transportation network ($\langle k \rangle$ = 7.8, $N$ = 3618) (32), on which service disruption cascades may occur (33); (iii) the Internet in 1999 ($\langle k \rangle$ = 3.8, $N$ = 3216) (34), where one may want to detect infected machines in a partially known bot network (35–37); and (iv) an email communication network ($\langle k \rangle$ = 9.6, $N$ = 1133) (5), through which sensitive information can spread. All networks are treated as undirected and unweighted. We perform a similar analysis on model networks that have realistic features (see the Supplementary Materials).

Figure 2 shows the performance of our method in detecting synthetic perturbations generated by the first version of the SI model, as compared to the results obtained using shortest paths (18, 19) (that is, the parsimony principle; see the Supplementary Materials) and label propagation (23, 24) (see the Supplementary Materials) for homogeneous perturbations. The reported values of $c$ are $c = 1/(\langle k \rangle + 0.05)$ and $c = 0.5$, two limiting values that ensure that perturbation propagates to a finite size of the network but does not reach percolation. The accuracy of a $k$–nearest neighbor classifier (27) is also evaluated in the Supplementary Materials. As expected, the performance of all inference methods generally decreases when the size of $\mathcal{O}$ decreases because less information is available. Performance also decreases with perturbation size (controlled by parameter $c$ in the perturbation generation process) and when the average connectivity of the network grows because the number of plausible propagation paths also grows in these cases (see fig. S2 for other topological features).

For the real networks we consider, our method outperforms shortest paths in all cases without exception (Fig. 2, A to D). Notably, when less than half of the system is observed, our approach also yields better accuracy than label propagation, with no exception. This result is particularly relevant because sparse observations are the most relevant in practice. For the Internet and the air transportation networks, our model outperforms label propagation even in a wider range of observed nodes.

Despite the fact that accuracy decreases for dense networks (Fig. 2D), we find that, for most cases, the AUC of our method is above 0.7 even for very small observations $\mathcal{O}$ (10% of nodes observed), large perturbations ($c$ close to 0.5), and relatively dense networks, such as the air transportation network.

The results for the SI model with heterogeneous probability of propagation are shown in Fig. 3. Also in this case, we find that our method outperforms label propagation in detecting the perturbation in most of the conditions, especially when the observed set is particularly small ($|\mathcal{O}| = 0.2$) and the perturbation is more homogeneously spread ($\delta_c \lesssim 0.5$).

**Inference of real metabolic perturbations**

We now consider the propagation of a real perturbation of human metabolism caused by the ingestion of a plant extract, a situation akin to what we described earlier (Fig. 4). In particular, we administered a plant extract to healthy volunteers and monitored the concentration of 188 metabolites before and after the administration of the extract (see the Supplementary Materials). We observe that, contrary to what one may expect, perturbed metabolites are spread over the whole human metabolic network [we reconstruct this network from the Kyoto Encyclopedia of Genes and Genomes (KEGG) database (*30*, *31*, *38*); see Materials and Methods] rather than localized into one particular metabolic pathway (Fig. 4A).

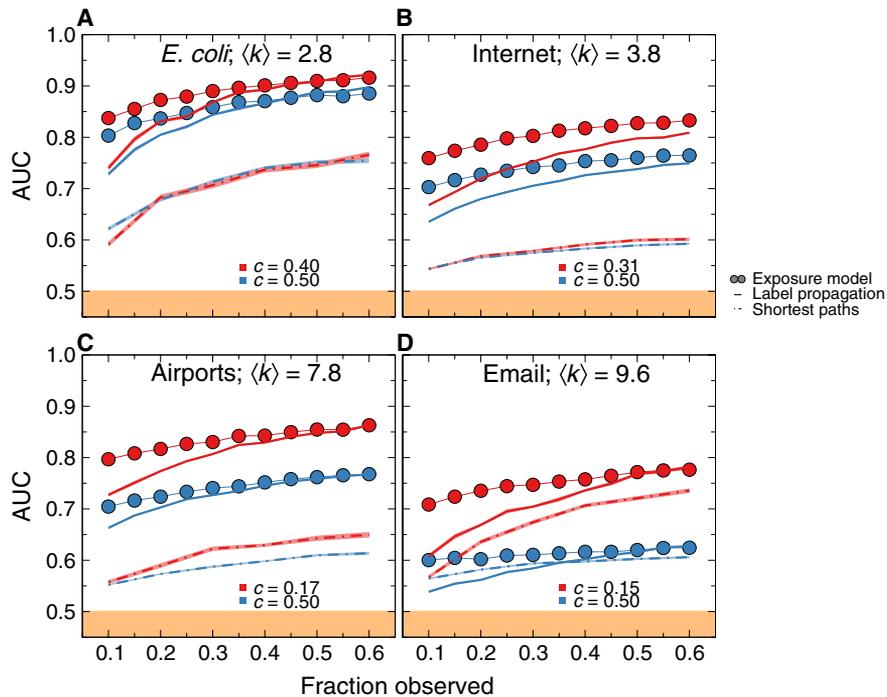

**Fig. 2. Accuracy of the node state inference for simulated perturbations with homogeneous perturbation probability $c$.** We plot the AUC (see Materials and Methods) as a function of the fraction of nodes whose state is observed, for the real networks described in the text. We compare the results of our algorithm on the basis of the exposure model (filled circles) with those obtained using shortest paths (dashed line) and label propagation (solid line). All points are averages over 100 perturbation repetitions (error bars are smaller than the markers). Different colors correspond to different values of the probability $c$ of contagion for the perturbation (the larger the value of $c$, the larger the perturbation). (**A**) Metabolic network of the bacterium *E. coli* (*4*, *30*, *31*), (**B**) reconstruction of the 1999 Internet at the autonomous system level (*34*), (**C**) global air transportation network (*32*), and (**D**) email communication network (*5*). Note that because observed nodes are picked at random, the observation is unbiased (see fig. S3); nonetheless, the performance of our method is not greatly affected if the observation is biased toward perturbed nodes (see fig. S4).

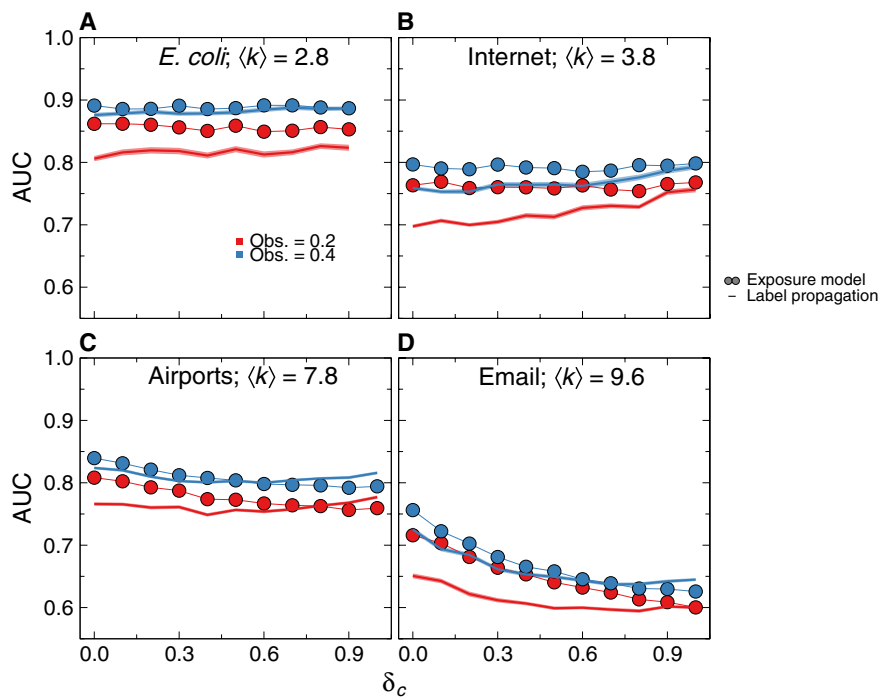

**Fig. 3. Accuracy of the node state inference for simulated perturbations with heterogeneous perturbation probability.** We fix the fraction of observed nodes and plot the AUC as a function of the range $\delta_c$ allowed for the probability of propagation $c$, for synthetic perturbations generated with a heterogeneous SI model (described more in detail in the Supplementary Materials). We show here the accuracy of the exposure model (filled circles) and of label propagation (solid line). Each point is an average over 100 different perturbations from distinct root nodes $\mathcal{R}$; error bars are smaller than the markers.

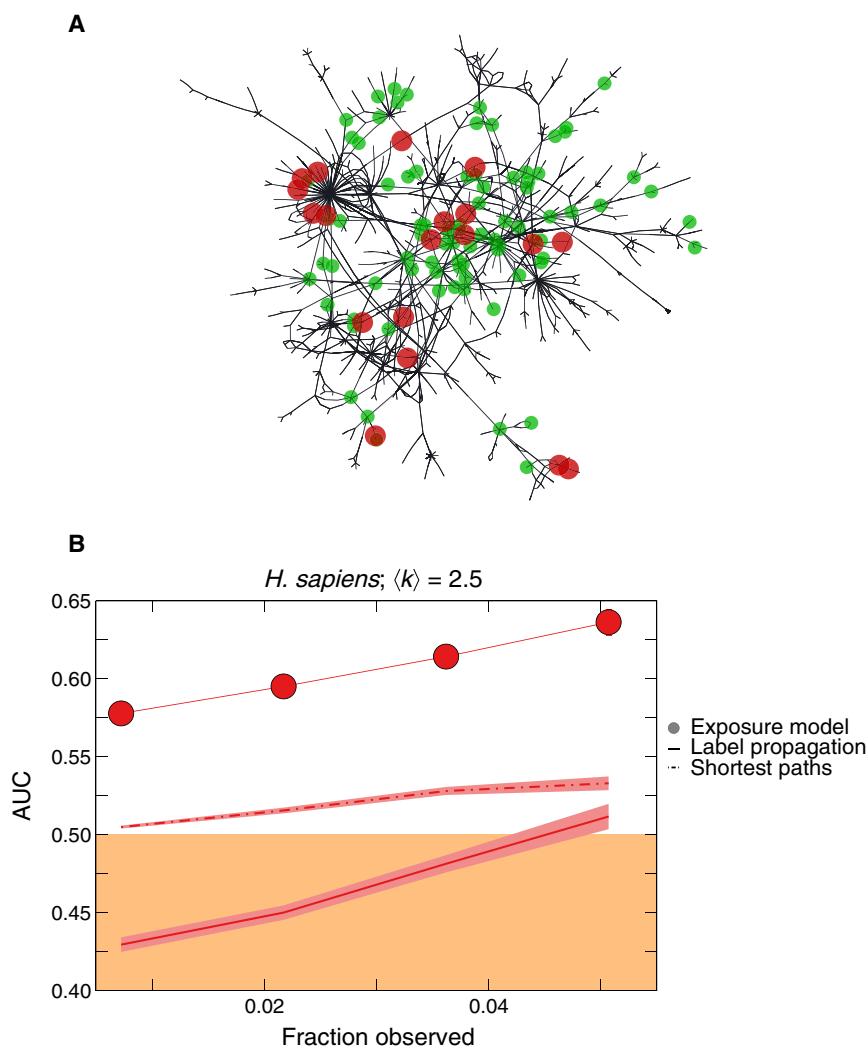

**Fig. 4. Results on human metabolism.** (**A**) The giant connected component of the human metabolic network (*30*, *31*), which we use to infer the real metabolic perturbation, as explained in Materials and Methods. Each node of the network is a metabolite; nodes are connected if a metabolic reaction processes both of them. For the sake of clarity, nodes are not shown. Large red circles indicate metabolites that changed concentration after the perturbation (observed perturbed nodes). Large green circles show metabolites whose concentration was not affected by the perturbation (observed unperturbed nodes). The observed nodes are just a minor fraction scattered all around the underlying network. (**B**) Upon hiding the state of a fraction of the observed nodes, we used the remaining set to infer the state of all the nodes. The graph shows the AUC (see Materials and Methods) obtained when using the exposure model (filled circles; error bars indicate 1 SE), shortest paths (dashed line; the shaded area indicates 1 SE), and label propagation (solid line; the shaded area indicates 1 SE) as a function of the fraction of nodes actually fed to the algorithm.

Because the complete propagation path is unknown in this case, we analyze whether different approaches are capable of identifying subsets of perturbed nodes (as we did in Fig. 2). Therefore, to assess the accuracy of each approach, we hide a subset of the observed metabolites (as if they were actually unobserved) and evaluate the AUC of the inference of the state of these hidden nodes.

Although, as we have mentioned, the perturbed nodes are scattered all over the network and the fraction of observed nodes is very low (1 to 5% of the whole network), with the exposure model, we achieve an AUC of 0.64, significantly above the AUC of both shortest paths and label propagation approaches, respectively. Note that the performance of LP is <0.5 for these small observations. This is because, whereas the perturbation is mostly localized along the central (denser) part of the network, where there is little information, label propagation tends to set periferic areas of unobserved nodes to the state of the closest node, which is typically perturbed. This results in typically labeling nonpertubed nodes as perturbed, therefore yielding an AUC below 0.5. Our results are notable considering the crude simplifications of the approach, namely, the assumption that the perturbation propagates exclusively through metabolic routes (when probably there are signaling pathways beyond metabolism) and the assumption that the network reconstruction is complete and accurate (*39*, *40*).

## DISCUSSION

The identification of perturbed nodes in sparsely observed complex systems is a central issue in many different fields, ranging from biology and medicine to the social sciences. Until now, approaches to the problem in network science had been limited to using the shortest paths approach, the so-called parsimony principle. In computer science, a few algorithms (for example, label propagation) were devised to deal with related problems. However, these algorithms had not been specifically applied to the problem of inferring perturbations on networks (therefore, they had not been compared to a method aimed at completing this task).

Here, we have demonstrated that a simple probabilistic model, which we call the exposure model, is appropriate for inferring perturbation propagation paths given very sparse information in synthetic and real networks and in synthetic and real perturbations. Our approach systematically outperforms an inference protocol based on shortest paths. This is probably because the shortest paths approach disregards any information provided by observed unperturbed nodes that do not fall in shortest paths between perturbed ones, whereas the exposure model allows for information to flow and is, in this sense, global rather than local. Additionally, the method we propose regularly performs better than inferring perturbations via label propagation when less than half of the system is observed. This situation is arguably expected to be met in most practical applications of our method. The observed performance gap may be explained, noticing that, for sparse observations, label propagation is prone to set the state of whole regions to that of any observed node in the neighborhood, whereas probabilities decay with distance in the exposure model.

We find that our method is more accurate for the inference of perturbation paths on real-world networks than on their synthetic counterparts (see the Supplementary Materials), showing that the assumptions behind our method are well suited to address real-world problems. This fact and the linear scaling of our algorithm with the number of links in the network make our approach very appealing to deal with real networks that contain hundreds of thousands of nodes.

Finally, our method allows us to infer a real metabolic perturbation on an actual metabolic network. The accuracy of our results is especially notable if we consider that there are many factors that we cannot take into account, such as regulatory mechanisms and enzymatic activities, not to mention the incompleteness of the metabolic reconstruction. Our results thus suggest that beyond being able to infer perturbation paths in a number of contexts, the framework we report can also be used as a tool to guide future hypothesis building and discovery, by, for instance, aiding in the identification of ramification points in the perturbation paths.

## MATERIALS AND METHODS

### Definition of the exposure model

In the exposure model, we assumed that perturbations propagate by proximity. Thus, we assumed (Eq. 1) that an unobserved node $i$ is perturbed with probability $\eta$ if it is exposed to the perturbation, that is, if at least one of $i$'s neighbors is perturbed; otherwise, the node is unperturbed with probability 1. We represented the state of the neighbors of $i$ as $\gamma_i$, which can be expressed in terms of the states $\{\sigma_j\}$ of each one of the neighbors of $i$ as follows

$$\gamma_i = 1 - \prod_{j \in \partial i}(1 - \sigma_j) \quad (7)$$

where $\partial i$ denotes the neighborhood of $i$. Then, for a fixed value of $\eta$, we can write the probability of node $i$ being in state $\sigma_j$ given $\gamma_i$ as follows

$$P(\sigma_i \mid \gamma_i, \boldsymbol{\sigma}^{\mathcal{O}}) = \begin{cases} \eta^{\sigma_i}(1-\eta)^{1-\sigma_i} & \text{if } \gamma_i = 1 \\ 1 - \sigma_i & \text{if } \gamma_i = 0 \end{cases} \quad (8)$$

### Belief propagation equations

The probability for node $i$ to be unperturbed for a fixed value of $\eta$ (for now, we drop the reference to $\eta$ as an argument for the sake of clarity; we focus on this parameter in the next section) is obtained by relating the posterior marginal of $i$ to that of its neighbors

$$P(\sigma_i | \boldsymbol{\sigma}^{\mathcal{O}}) = \sum_{\boldsymbol{\sigma}_{\setminus i}} P(\boldsymbol{\sigma}|\boldsymbol{\sigma}^{\mathcal{O}}) = \sum_{\boldsymbol{\sigma}_{\partial i}} P(\sigma_i | \gamma_i, \boldsymbol{\sigma}^{\mathcal{O}}) P(\boldsymbol{\sigma}_{\partial i} | \boldsymbol{\sigma}^{\mathcal{O}}) \quad (9)$$

where $\boldsymbol{\sigma}_{\setminus i}$ is the set of unobserved nodes except $i$, $\boldsymbol{\sigma}_{\partial i}$ are the neighbors of $i$, and $P(\sigma_i | \gamma_i, \boldsymbol{\sigma}^{\mathcal{O}}) \equiv P(\sigma_i | \boldsymbol{\sigma}_{\partial i}, \boldsymbol{\sigma}^{\mathcal{O}})$. We then approximated the joint posterior $P(\boldsymbol{\sigma}_{\partial i} | \boldsymbol{\sigma}^{\mathcal{O}})$ as $P^{(i)}(\boldsymbol{\sigma}_{\partial i} | \boldsymbol{\sigma}^{\mathcal{O}})$ by calculating it in the absence of node $i$. Furthermore, we assumed that the system is tree-like, such that, without $i$, this joint posterior factorizes, that is, $P^{(i)}(\boldsymbol{\sigma}_{\partial i}|\boldsymbol{\sigma}^{\mathcal{O}}) = \prod_{k \in \partial i} P^{(i)}(\sigma_k|\boldsymbol{\sigma}^{\mathcal{O}})$. By doing so, one can write $P(\sigma_i|\boldsymbol{\sigma}^{\mathcal{O}})$, as in Eq. 3. The self-consistent equations for the marginals $P^{(i)}(\sigma_k|\boldsymbol{\sigma}^{\mathcal{O}})$ are derived by removing node $i$ from the system

$$P^{(i)}(\sigma_k|\boldsymbol{\sigma}^{\mathcal{O}}) = \sum_{\boldsymbol{\sigma}_{\partial k \setminus i}} P(\sigma_k|\gamma_k^{(i)}, \boldsymbol{\sigma}^{\mathcal{O}}) \prod_{j \in \partial k \setminus i} P^{(k)}(\sigma_j|\boldsymbol{\sigma}^{\mathcal{O}}) \quad (10)$$

where, independently of the state of $i$, $\gamma_k^{(i)}$ is equal to 0 if none of the remaining neighbors of $k$ is perturbed or is equal to 1 if at least one of them is perturbed. The parameterization in terms of the beliefs $\psi_i$ and messages $\psi_i^{(k)}$ is obtained by noting that $\sigma_i$ is a binary variable and $P^{(k)}(\sigma_i = 0|\boldsymbol{\sigma}^{\mathcal{O}})$ fully specifies each marginal. Expressing $P(\sigma_i = 0|\gamma_i, \boldsymbol{\sigma}^{\mathcal{O}})$ according to Eq. 8 and plugging this expression into Eq. 3, we obtain the expressions for $\psi_i^{(k)}$ and $\psi_i$ given in Eqs. 4 and 5. The resulting set of equations may be solved iteratively.

### Expectation maximization and inference algorithm

To compute the probability marginals for each $\sigma_i$, one would need to integrate over all values of the unobserved parameter $\eta$ that controls how efficiently the perturbation propagates (Eq. 2). Rather than exactly evaluating the integral, we used a saddle point approximation by computing the $\eta^\star$ value that maximizes the posterior $P(\eta|\boldsymbol{\sigma}^{\mathcal{O}})$. To do so, we maximized the log-weight $\log \Sigma_{\boldsymbol{\sigma}} P(\eta|\boldsymbol{\sigma}^{\mathcal{O}})$ (note that because of the dependence on $\eta$, the trace is not normalized). In particular, we have

$$\frac{\partial}{\partial \eta} \log \sum_{\boldsymbol{\sigma}} P(\boldsymbol{\sigma}, \eta | \boldsymbol{\sigma}^{\mathcal{O}}) = \frac{\partial}{\partial \eta} \frac{1}{N} \sum_i \log \sum_{\sigma_i} P(\sigma_i, \eta | \boldsymbol{\sigma}^{\mathcal{O}})$$

$$= \frac{1}{NZ} \sum_i \sum_{\sigma_i, \sigma_{\partial i}} \left( \gamma_i \frac{\sigma}{\eta} - \gamma_i \frac{1-\sigma}{1-\eta} \right) P(\sigma_i, \eta, \gamma_i | \boldsymbol{\sigma}^{\mathcal{O}}) \quad (11)$$

$$= \frac{1}{N} \sum_i \left( \frac{\langle \gamma_i \sigma_i \rangle}{\eta} - \frac{\langle \gamma_i (1-\sigma_i) \rangle}{1-\eta} \right)$$

where $Z = \sum_{\sigma_i} P(\sigma_i, \eta | \boldsymbol{\sigma}^{\mathcal{O}})$ and, hence, $\langle \ldots \rangle$ is a properly normalized average. To differentiate with respect to $\eta$, we used here the definitions of $P(\boldsymbol{\sigma}, \eta)$ in Eq. 8, and we noticed that because $P(\sigma_i) = \sum_{\boldsymbol{\sigma}_{\setminus i}} P(\boldsymbol{\sigma})$, one has $\sum_{\sigma_i} P(\sigma_i) = \sum_{\boldsymbol{\sigma}} P(\boldsymbol{\sigma})$. We thus performed the trick of carrying out this normalization $N$ times [over each node $i$, $(\sum_{\boldsymbol{\sigma}} P(\boldsymbol{\sigma}))^N$] and normalized over $N$ accordingly.

Equating Eq. 11 to zero yields the $\eta^\star$ value that maximizes the weight

$$\eta^\star = \frac{\mathcal{N}_{11}}{\mathcal{N}_{01} + \mathcal{N}_{11}} \quad (12)$$

where $\mathcal{N}_{11} \equiv 1/N \sum_i \langle \gamma_i \sigma_i \rangle$ and $\mathcal{N}_{01} \equiv 1/N \sum_i \langle \gamma_i (1 - \sigma_i) \rangle$ are the expected number of perturbed and unperturbed nodes with perturbed neighbors, respectively. These quantities could, in principle, be obtained from the node states $\sigma_i$. However, within our belief propagation approach, we only have the estimates of the marginals $\{\psi_i\}$ and $\{\psi_i^{(k)}\}$. Therefore, we wrote $\eta^\star$ in terms of these marginals, as shown in Eq. 6, and numerically solved Eqs. 4 to 6 using the algorithm outlined in the Supplementary Materials.

Note that $\eta^\star$ is an effective parameter that does not necessarily bear resemblance to any of the real parameters of the perturbation. However, for the SI type of perturbations we generated on the real networks we studied (shown in Fig. 2), we found that the values of $\eta^\star$ and the propagation probability $c$ are very close (see fig. S5).

### Algorithm accuracy

Given the probability of each unobserved node being perturbed, we assessed the performance of our approach by computing the AUC (29). In practice, we ranked all unobserved nodes from smallest to largest $\psi_i$ and computed the frequency with which a true perturbed node is ranked above a true unperturbed node (following the usual convention, cases in which a perturbed and an unperturbed node are assigned the same $\psi_i$ are considered half correct). The AUC is a real number ranging from 0 to 1—the closer to 1, the more true perturbed nodes are ranked above true unperturbed ones. Conversely, an AUC value of 0.5 indicates pure randomness in performing the ranking, in which case the inference protocol is no better than a coin toss.

### Human metabolic perturbation

A metabolomic analysis of human blood plasma (see Supplementary Materials) was carried out on healthy volunteers to obtain a list of experimentally perturbed and unperturbed metabolites after ingestion of a plant extract.

We used the observed perturbed and unperturbed metabolites to infer the state of all remaining metabolites included in the reconstruction of human metabolism obtained from the KEGG database (30, 31). For the reconstruction, we considered only links between main reactant pairs in each reaction (4, 30, 31). This reconstruction features 1422 metabolites connected by 1760 reactions. We considered this reconstruction to be the underlying network on which perturbation propagates. Note that 85 of the 188 metabolites identified by the experimental analysis were not present in the reconstruction; thus, our observation $\mathcal{O}$ was reduced to a pool of only 103 metabolites (19 perturbed and 84 unperturbed).

To evaluate the performance of the exposure model in detecting the perturbation, we hid a fraction of the set $\mathcal{O}$ and measured the AUC for the inferred nodes in the hidden subset of $\mathcal{O}$. Figure 4B shows how the exposure algorithm performs better than the shortest path algorithm, which is especially notable, taking into account that observed nodes are scattered throughout the network (see Fig. 4A). Note that the AUC values we obtained seemed to be lower than those reported for synthetic perturbations; this is due to the fact that the observation amounts to a mere 7% of the total number of metabolites and would therefore correspond to very low values in the $x$ axis of the plots in Fig. 2, in which AUC values are comparable to the ones we report in Fig. 4.

## SUPPLEMENTARY MATERIALS

Supplementary material for this article is available at http://advances.sciencemag.org/cgi/content/full/2/10/e1501638/DC1
Supplementary Materials and Methods
fig. S1. Belief propagation method used in our approach.
fig. S2. Topological effects on the accuracy of the model.
fig. S3. Dependence of the algorithm performance on perturbation size and fraction of observed nodes.
fig. S4. Dependence of the performance of our approach versus the fraction of perturbed nodes in the observation.
fig. S5. Expectation maximization-estimated model parameter $\eta^\star$ versus perturbation parameter $c$.
fig. S6. Accuracy of a $k$-neighbor classifier at detecting perturbations.
fig. S7. Perturbations on synthetic networks.
References (42–44)

**Acknowledgments:** We thank A. Aguilar-Mogas, A. Arenas, A. Gavaldà-Miralles, A. Godoy-Lorite, O. Senan-Campos, and T. Vallès-Català for their comments and suggestions. **Funding:** This work was supported by a James S. McDonnell Foundation Research Award, grants PIRG-GA-2010-277166 (R.G.) and PIRG-GA-2010-268342 (M.S.-P.) from the European Union, and grants FIS2013-47532-C3 and FIS2015-71563-ERC from the Spanish Ministerio de Economía y Competitividad. J.J. is supported by grants from the Plan Nacional de I+D+I, Spain, and Instituto de Salud Carlos III [grant PI15/00285, cofounded by the European Regional Development Fund (FEDER)]. **Author contributions:** F.A.M., M.S.-P., and R.G. designed the research; F.A.M., J.W., and R.G. performed the research; F.A.M. and M.S.-P. analyzed the data; F.A.M., M.S.-P., and R.G. wrote the manuscript; R.B.-D. and J.J. contributed data and materials. **Competing interests:** The authors declare that they have no competing interests. **Data and materials availability:** All data needed to evaluate the conclusions in the paper are present in the paper and/or the Supplementary Materials. Additional data related to this paper may be requested from the authors.

Submitted 13 November 2015
Accepted 22 September 2016
Published 21 October 2016
10.1126/sciadv.1501638

**Citation:** F. A. Massucci, J. Wheeler, R. Beltrán-Debón, J. Joven, M. Sales-Pardo, R. Guimerà, Inferring propagation paths for sparsely observed perturbations on complex networks. *Sci. Adv.* **2**, e1501638 (2016).

**Inferring propagation paths for sparsely observed perturbations on complex networks**

Francesco Alessandro Massucci, Jonathan Wheeler, Raúl Beltrán-Debón, Jorge Joven, Marta Sales-Pardo and Roger Guimerà